\documentclass[3p,10pt,sort&compress, colorlinks,linkcolor=blue, citecolor=blue, urlcolor=blue]{article}

\usepackage[utf8]{inputenc}
\usepackage{tabularx}
\usepackage{hyperref}

\usepackage[utf8]{inputenc}

\usepackage{amssymb}
\usepackage{amsmath,mathrsfs}
\usepackage{empheq}

\usepackage[sc]{mathpazo}
\usepackage{bm}

\usepackage{color,soul}

\usepackage{extarrows}

\usepackage{verbatim}

\usepackage{lineno}

\usepackage{siunitx}

\usepackage{stfloats}

\usepackage{bm}

\usepackage{graphicx}

\usepackage{caption}
\usepackage{subcaption}

\usepackage{multicol}

\usepackage{multirow}

\usepackage{mhchem}

\usepackage[toc,page]{appendix}

\usepackage{bm}

\usepackage{float}

\usepackage{enumerate}

\usepackage{subfiles}

\usepackage{threeparttable}

\usepackage{tabularx}
\usepackage{booktabs}
\usepackage[table,xcdraw]{xcolor}
\usepackage{makecell}

\usepackage{sectsty}
\usepackage[toc,page]{appendix}
\renewcommand{\eqref}[1]{Eq.~$($\ref{#1}$)$}

\def\stress{\boldsymbol{\sigma}}
\def\strain{\boldsymbol{\varepsilon}}

\def\Hv{\mathbf{H}}
\def\Ev{\mathbf{E}}
\def\Dv{\mathbf{D}}
\def\Bv{\mathbf{B}}
\def\jv{\mathbf{j}}

\def\rv{\mathbf{r}}
\def\uv{\mathbf{u}}
\usepackage{authblk}

\providecommand{\keywords}[1]
{
	\small	
	\textbf{\textit{Keywords---}} #1
}

\begin{document}

    \title{What can machine learning help with microstructure-informed materials modeling and design?}
	\author[*]{Xiang-Long Peng*, Mozhdeh Fathidoost, Binbin Lin, Yangyiwei Yang, Bai-Xiang Xu}
	
	\affil{\small Mechanics of Functional Materials Division, Institute of Materials Science, Technische Universit\"at Darmstadt, Darmstadt 64287, Germany}

    \affil[*]{Corresponding authors: \url{xianglong.peng@tu-darmstadt.de} (Xiang-Long Peng); \url{xu@mfm.tu-darmstadt.de} (Bai-Xiang Xu)}
	\date{}
	\maketitle
	\renewcommand\Authands{ and }		
\begin{abstract}
Machine learning techniques have been widely employed as effective tools in addressing various engineering challenges in recent years, particularly for the challenging task of microstructure-informed materials modeling. This work provides a comprehensive review of the current machine learning-assisted and data-driven advancements in this field, including microstructure characterization and reconstruction, multiscale simulation, correlations among process, microstructure, and properties, as well as microstructure optimization and inverse design. It outlines the achievements of existing research through best practices and suggests potential avenues for future investigations. Moreover, it prepares the readers with educative instructions of basic knowledge and an overview on machine learning, microstructure descriptors and machine learning-assisted material modeling, lowering the interdisciplinary hurdles. It should help to stimulate and attract more research attention to the rapidly growing field of machine learning-based modeling and design of microstructured materials.
\end{abstract}
\section{Introduction}
Microstructured materials manifest in diverse engineering scenarios in forms of polycrystalline microstructures, inclusion-matrix composites, bicontinuous composites, porous structures \cite{bargmann2018generation}, etc. Their macroscopic properties strongly depend on the underlying microstructural features. This presents an avenue for achieving tailored macroscopic properties which are unattainable in the base materials. Thus, comprehending the quantitative impact of microstructures on macroscopic properties is desired.
The fabrication or synthesis of microstructured materials is a complex process subjected to various process parameters. These parameters significantly affect the resultant microstructures, and thereby the corresponding macroscopic properties. Thus, the design, fabrication, and application of microstructured materials necessitate a deep understanding of the interplay between process and microstructure, microstructure and property, and, ultimately, direct process-property relations. Traditional experimental and numerical methods, though valuable, cannot efficiently tackle these challenging tasks.

In recent years, artificial intelligence and machine learning (ML) have emerged as transformative forces, driving profound economic and social changes. They have become pivotal technologies in various research domains, including material science and engineering \cite{himanen2019Review}. The integration of ML and data-driven techniques into scientific research methodologies has given rise to what is known as the fourth research paradigm \cite{hey2009fourth}, extending the third paradigm of computational science.
This paradigm shift signifies a departure from traditional scientific approaches, embracing the power of data-driven insights and predictive capabilities facilitated by ML. A number of review papers have illuminated the remarkable progress achieved through the application of ML methods in different scenarios in the broad field of material sciences and mechanics. These include materials design \cite{yang2019establishing,dimiduk2018review,wang2022datareview,jin2022intelligent,lee2023data}, atomistic simulations \cite{himanen2019Review,vasudevan2019review}, multiscale modeling and simulation \cite{bishara2023}, mechanics of materials \cite{bock2019review,kumar2022machine,dornheim2023neural}, etc. These reviews collectively demonstrate the efficiency and effectiveness of ML methods in addressing a spectrum of challenging problems in material design, modeling, and engineering.

As summarized in Table~\ref{tab1}, ML can assist materials modeling in terms of constitutive surrogates, energy functional surrogates or field predictors, in alignment with continuum mechanics theory. However, the existing research and literature articles are mostly dedicated to homogeneous material or homogenized microstructure. In fact, for these cases, traditional continuum models have reached a high level of sophistication. The ML-based material models frequently emerge only as alternatives or supplements to these established frameworks. In contrast, there are hardly continuum models which take microstructure descriptors or local properties as direct variables, mainly due to the fact that the intricate and diverse nature of microstructures and the mostly unknown material physics implications of microstructures exceed the capabilities of conventional methodologies. In this sense, microstructure-informed materials modeling presents an ideal opportunity for ML-based modeling approaches to flourish. 

\begin{table}[]
\small
\caption{Classification of ML-based material models. Abbreviations: "temp.", "chem.", "pot.", and "thermody." denote "temperature", "chemical", "potential", and "thermodynamically", respectively.}
\begin{threeparttable}[b]
\begin{tabular}{@{}ccccc@{}}
\toprule
  & \multicolumn{2}{c}{Homogeneous Material} & \multicolumn{2}{c}{Heterogeneous Material} \\ \cmidrule(lr){2-3} \cmidrule(lr){4-5}
\multirow{-2}{*}{\makecell{ML-based\\Models}} & Input X   & Output Y  & Input X     & Output Y    \\ \cmidrule(r){1-5}
\makecell[c]{I. \\Constitutive
\\surrogate\\e.g.,~\cite{LOGARZO2021,asheri2023}}
& \makecell[l]{\textbf{Physical Fields:}\\ e.g., strain $\strain$, $\strain^\mathrm{v}$, $\strain^\mathrm{p}$, \\
    electromagnetic  \\field $\Ev$, $\Hv$ \\
    \textbf{Driving Forces:}\\
    e.g., temp. gradient $\nabla T$, \\
    chem. pot. gradient $\nabla \mu$}   & 
    
\makecell[l]{\textbf{Thermody. Conjugated}\\\textbf{Quantities}:\\ e.g., stress $\stress$, \\
    electromagnetic\\ inductions $\Dv$, $\Bv$ \\
    \textbf{Kinetic Response:}\\
    e.g., heat flux $\jv_q$, \\
    chem. species flux $\jv_c$}
& \makecell[l]{\textbf{Physical Fields}:\\ e.g., strain $\strain$, $\strain^\mathrm{v}$, $\strain^\mathrm{p}$, \\
    electromagnetic  \\field $\Ev$, $\Hv$ \\
    \textbf{Driving Forces:}\\
    e.g., temp. gradient $\nabla T$,\\
    chem. pot. gradient $\nabla \mu$
\\ \textbf{+} \\ \textbf{Microstructural} \\ \textbf{Descriptors}}    &
\makecell[l]{\textbf{Thermody. Conjugated}\\\textbf{Quantities}:\\ e.g., stress $\stress$, \\
    electromagnetic\\ inductions $\Dv$, $\Bv$ \\
    \textbf{Kinetic Response:}\\
    e.g., heat flux $\jv_q$, \\
    chem. species flux $\jv_c$}                          \\ \hline
\makecell[c]{II. \\Energy functional \\ surrogate\\
e.g.,~\cite{fernandez2022material,kalina2023fe}}  & \makecell[l]{\textbf{Tensor Invariant}:\\ e.g., strain invariant\\ $I_{\strain}$,... $I_{\strain^\mathrm{v}}$,... $I_{\strain^\mathrm{p}}$}& \makecell[l]{\textbf{Energy/Thermody.}\\ \textbf{Potentials}:
\\ e.g., dissipated energy $U$,
\\ free energy density\\ $f$, $g$}
&  \makecell[l]{\textbf{Tensor Invariant}:\\ e.g., strain invariant\\ $I_{\strain}$,... $I_{\strain^\mathrm{v}}$,... $I_{\strain^\mathrm{p}}$
\\ \textbf{+} \\ \textbf{Microstructural} \\ \textbf{Descriptors}}

& \makecell[l]{\textbf{Energy/Thermody.  }\\\textbf{Potentials}:
\\ e.g., dissipated energy $U$,
\\ free energy density\\ $f$, $g$}                      \\ \hline
\makecell[c]{III. \\Field predictor\\e.g.,~\cite{REZAEI2022115616,yang2019establishing,lin2023machine} } & \makecell[l]{\textbf{Coordinates} $\rv$}  & \makecell[l]{\textbf{Solution Fields}\\
e.g., displacement $\uv(\rv)$, \\
stress $\stress(\rv)$, \\
temperature $T(\rv)$,\\ 
electromagnetic \\
inductions $\Dv(\rv)$, $\Bv(\rv)$}  &
\makecell[l]{
\textbf{Coordinates} $\rv$ \\
\textbf{+} \\ \textbf{Local material properties}:
\\ e.g., mechanical properties $C(\rv)$,\\
thermal conductivity $\kappa(\rv)$,\\
permittivity $\xi(\rv)$
}&  
\makecell[l]{\textbf{Solution Fields}: \\
e.g., displacement $\uv(\rv)$, \\
stress $\stress(\rv)$, \\
temperature $T(\rv)$,\\ 
electromagnetic \\
inductions $\Dv(\rv)$, $\Bv(\rv)$}                  
\\ 
\bottomrule
\end{tabular}
\end{threeparttable}
\label{tab1}
\end{table}

\begin{figure}[p]
    \centering
\includegraphics[width=0.95\textwidth]{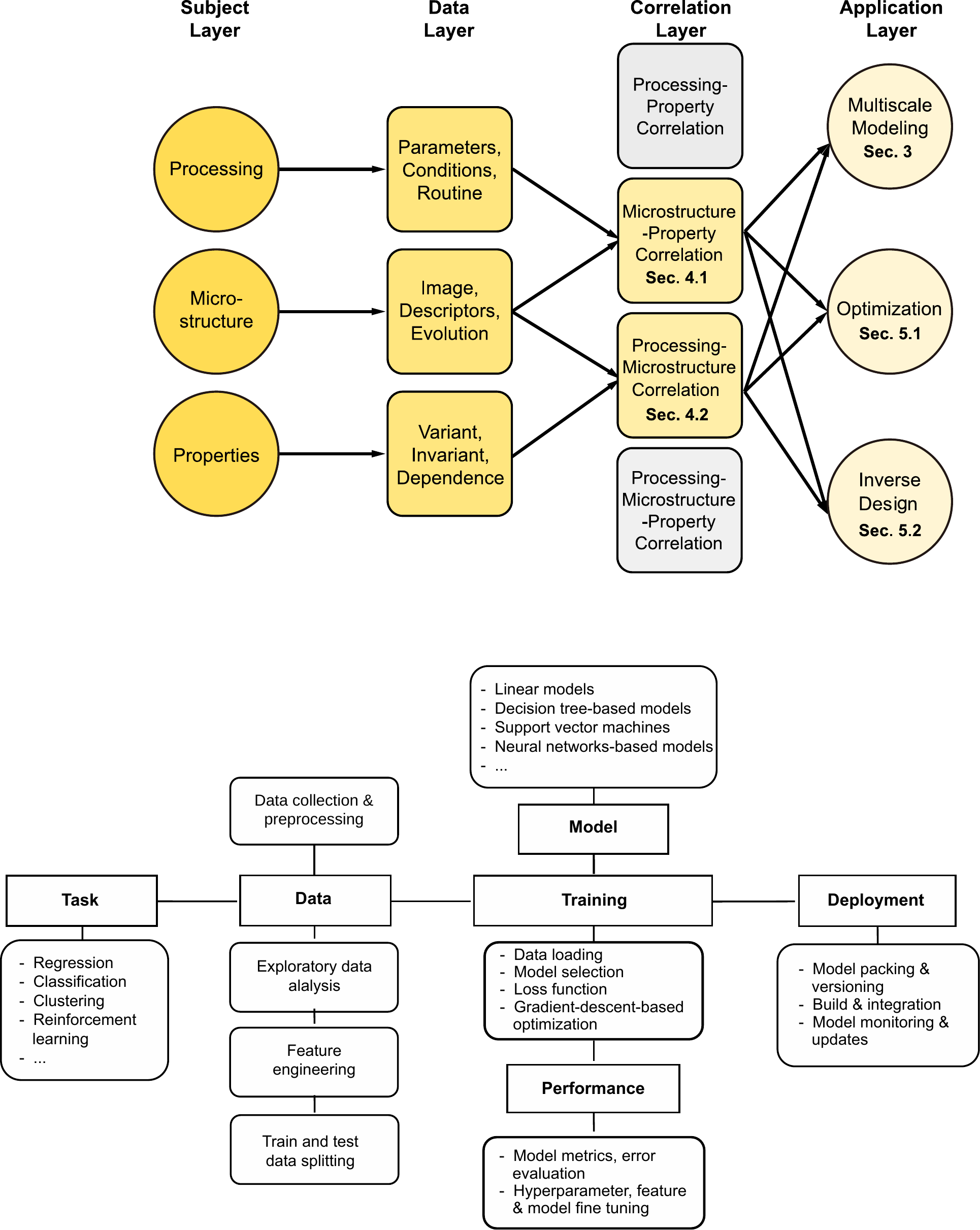}
    \caption{a) Schematic of different aspects and their connections relevant to ML-based modeling and design of microstructured materials. b) Illustration of the general workflow for ML.}
    \label{fig:machine learning}
\end{figure}
To promote this, we present a comprehensive review collectively addressing the roles that ML can play in different areas relevant to the modeling and design of microstructured materials including multiscale simulation, microstructure correlation and sensitivity analysis, and microstructure optimization and inverse design, as outlined in Fig.\,\ref{fig:machine learning}a. To this end, we highlight the representative works/best practices from the literature to demonstrate the potential of ML methods in solving challenging problems in the relevant fields. Based on the reviewed works and our research experiences and knowledge, we suggest the remaining challenges and urgent fields to which ML can significantly contribute, aiming at stimulating and guiding future investigations. Additionally, this work is valuable for instructive and educative purposes, empowering researchers with the knowledge and confidence needed to harness the potential of ML methods for advancing the modeling and design of materials. The paper is structured as follows.

Section~\ref{fundamental} introduces fundamentals on microstructure reconstruction and characterization and ML methods. Modeling and designing microstructured materials necessitate reconstructing microstructures and characterizing their features, which is already a well-developed field with significant progress. We will illustrate the relevant key points. Besides, a compact yet informative introduction on the workflow of ML methods including task definition, data collection and preprocessing, model selection, training, performance evaluation, and deployment is provided. These bases are included for educative and instructive purposes and are also necessary for the readers to understand the content of the current work and the reviewed literature. 

Section~\ref{Mult_Simu} is dedicated to data-driven multiscale simulation. The multiscale nature of microstructured materials necessitates a multiscale modeling and simulation strategy to characterize their properties and behaviors, which is a challenging task. On the one hand, characterizing and reconstructing microstructures (see Section~\ref{fundamental}) are non-trivial tasks, especially for sophistical microstructures where high dimensional and diverse features are involved. On the other hand, in a multiscale framework where the scale-separation is assumed, a scale-bridging scheme is crucial. At the macroscale, the object is effectively modeled as a homogeneous material at least locally. The corresponding macroscopic/effective material properties/laws are obtained through microscale simulations on representative volume elements (RVEs). Existing multiscale methods are generally classified as concurrent (e.g., $\text{FE}^2$ method) or sequential types. The concurrent method couples microscale and macroscale on-the-fly, ensuring high accuracy but is computationally expensive. The sequential framework employs a pre-constructed surrogate model for macroscale simulations, offering computational efficiency but relying on experiences and lacking reliability for complex problems. These drawbacks and limitations of traditional multiscale methods can be remedied by ML-based surrogate modeling. Although the ML method can be applied to homogeneous materials, we focus on those for inhomogeneous microstructured materials. How these ML-based material models are integrated into to both sequential and concurrent multiscale schemes will be discussed.

Section~\ref{Correlation_sensitivity} addresses how ML methods are employed to construct correlations among processes, microstructures, and properties. These correlations are crucial for tailoring microstructures to achieve specific macroscopic properties. However, the corresponding regression tasks typically involve highly complex and nonlinear relationships with diverse and high-dimensional inputs and outputs, posing challenges for conventional methodologies. In fact, due to their powerful fitting capability, ML models are well-suited for complex regression tasks. To unravel how process parameters influence microstructural features and, subsequently, how these features impact material properties, vast amounts of data are naturally required. This dataset can originate from experimental measurements or traditional and ML-aided numerical simulations. Subsequently, it is adopted to train ML surrogate models which act as efficient predictors. The description of a process or a microstructure may involve numerous parameters. However, not all of these parameters are equally important in the correlations under investigation. With the aid of the ML surrogate models, the sensitivity analysis can be conducted to identify the subset of parameters that significantly influence the correlations. 

Section~\ref{optimization_inverse} presents ML-based microstructure optimization and inverse design. They play a critical role in designing microstructured materials with on-demand properties. Microstructure optimization is to find a microstructure that results in extreme properties (e.g., minimal or maximum). The inverse design focuses on predicting a microstructure with specified target properties not necessarily to be the extreme values. Both tasks are extremely challenging due to their inherent high dimensionality, strong nonlinearity, and high computational cost. In this regard, ML methods if properly employed can efficiently solve these challenging tasks. In the optimization case, the ML method can be directly used as an optimization algorithm such as Bayesian optimization. Alternatively, incorporating the ML-based surrogate model into the traditional optimization methods as an efficient property predictor can largely promote efficiency. In the inverse design case, ML models can act as surrogate models directly correlating the effective properties with microstructural features.

Section~\ref{concluding} outlines the concluding remarks and perspectives. The scope of the current work is not only on reviewing existing literature but also aims at providing comprehensive perspectives and outlooks on the unexplored problems and potential opportunities for advancements in the field of ML-aided modeling and design of microstructured materials. Hopefully, it will inspire and guide future research endeavors in this fascinating field.

\section{Fundamentals}
\label{fundamental}
For educative and instructive purposes, we introduce some fundamentals on microstructure characterization and reconstruction and ML methods in this section, which are also essential for the readers to understand the content of the current work and the relevant literature. 

\subsection{Microstructure Characterization and Reconstruction}
\label{sec:Microstructure Characterization}
\begin{figure}[p]
    \centering
\includegraphics[width=0.7\textwidth]{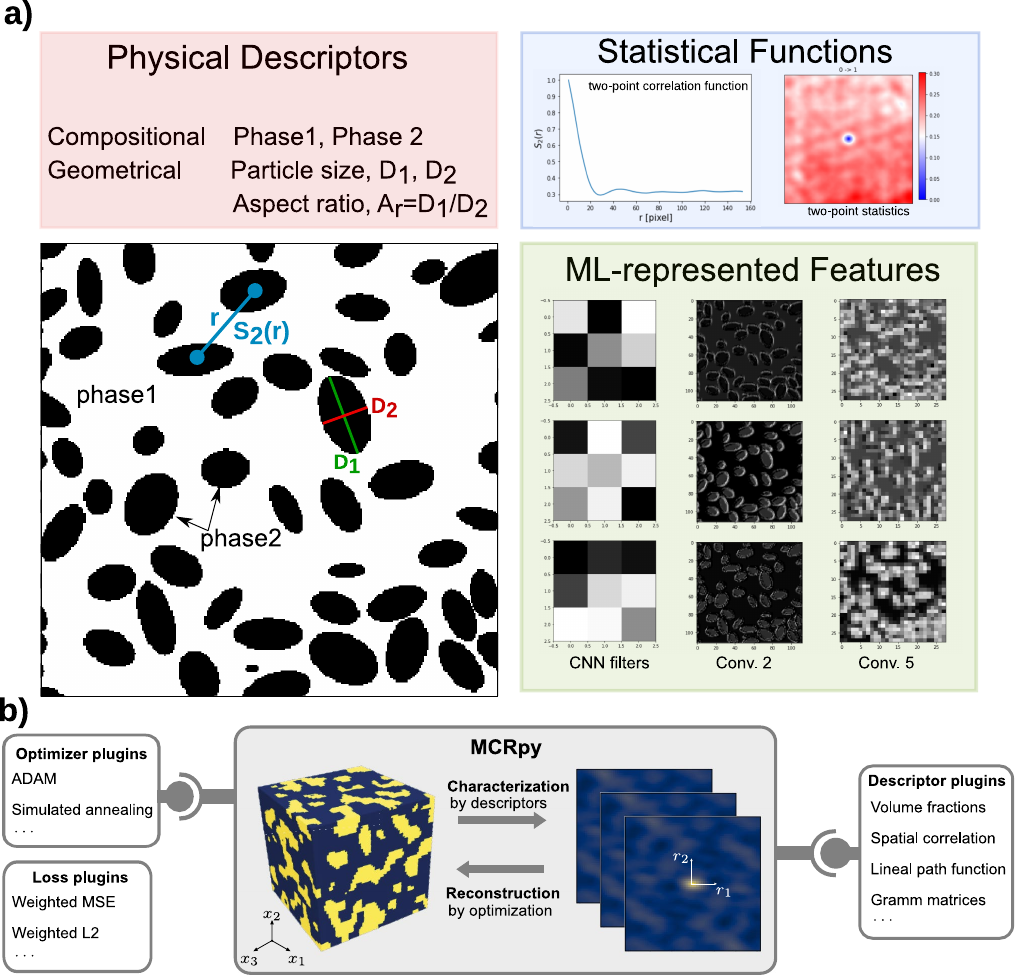} 
    \caption{Microstructure characterization and reconstruction. a) Characterization of a two-phase microstructure by three different types of representations, i.e., physical descriptors such as the particle size and aspect ratio, statistical descriptions based on the two-point correlation function, and ML-based representation illustrated by the feature maps learned through the convolutional layers of a CNN model and a few examples of the filters. b) The workflow of the Python package MCRpy for microstructure characterization and reconstruction~\cite{seibert2022microstructure}. Figure b) is adapted from~\cite{seibert2022microstructure} (published under license CC BY 4.0 \url{https://creativecommons.org/licenses/by/4.0.}).}
    \label{fig:microstructure reconstruction}
\end{figure}

Microstructure characterization and reconstruction is one of the key steps in the process of modeling and design of microstructured materials. On the one hand, investigating the microstructure-property correlation requires characterizing the simulated microstructures. On the other hand, the optimization and inverse design of microstructures necessitate the reconstruction of microstructures with predicted microstructural features.
The literature offers various ML methods to represent and characterize microstructures that can be integrated into computational and ML models for property prediction and materials design~\cite{bostanabad2018computational,Lin2022npjImage}. Details go beyond the scope of this review article. Instead, a short summary on various microstructure descriptors is given below, according to a broad classification:
\begin{itemize}
    \item Physical descriptors:
    Employing physical descriptors aims to characterize a microstructure by a limited set of features. 
    Physical descriptors are either deterministic or statistical and can be classified into geometric, compositional/chemical, or dispersional types. Composition descriptors identify different phases within a material and quantify their respective proportions, e.g., volume fractions and porosity. Dispersion status descriptors characterize the spatial arrangement of different phases and their proximity to neighboring elements, such as the nearest neighbor distance. Geometry descriptors focus on the geometrical details, such as the size distribution, surface area, and aspect ratio of inclusions or fibers in composites.  These descriptors possess clear physical meaning and are directly interpretable for users. However, representing microstructures with such a few descriptors may result in the loss of microstructure information. For this reason, a feature selection process based on, e.g. variable ranking approach and exploratory factor analysis~\cite{bostanabad2018computational} is necessary. 
\item Statistical function based descriptors:
Microstructures can be characterized by utilizing statistical correlation functions, commonly known as n-point correlation functions~\cite{torquato2005random} and their variants, e.g., lineal-path function, two-point cluster correlation function and spectral density function~\cite{bostanabad2018computational,xu2021guiding}. These statistical functions essentially quantify the spatial correlations among different points in a microstructure. This type of representation is suitable for reconstructing statistically equivalent microstructures~\cite{guo2014accurate,karsanina2015universal}. 
However, the lengthy and abstract statistical functions-based descriptors cannot be directly used for microstructure-property correlation. To remedy this issue, dimensionality reduction techniques such as principal component analysis (PCA) and linear/nonlinear embedding methods can be employed.

\item  ML-represented descriptors:
ML methods can be directly used for microstructure characterization. For instance, taking the digitalized microstructure images as inputs, convolutional neural network (CNN) models can extract microstructure features by convolutional layers with different filters, which are then converted into a low-dimensional vector by pooling and flattening layers. Once well-trained by a large microstructure data, CNN models are capable of characterizing the features of a system of microstructures. These extracted features can act as inputs for further supervised and/or unsupervised ML tasks such as classification and regression. Another type is the instance-based learning \cite{sundararaghavan2005classification}. It first creates a database that contains representative sample instances, then defines the feature space (e.g., the statistical functions) and similarity metrics applicable to all the samples in the database (e.g. hierarchical support vector machine classification model), and finally conducts either a brute force or algorithmic search to find the most similar instance in the database upon query. In this case, not only the database should be large enough but also the feasible space must be well covered. Since the size of the database increases exponentially, it is essential to develop a rapid searching mechanism that can efficiently locate the most similar instance in the database upon a query. 
  \end{itemize}
As an example, we illustrate how the above three methods can be exploited to represent the microstructural features of a two-phase matrix-inclusion composite in Fig.~\ref{fig:microstructure reconstruction}a. In this example, the two-phase microstructure was first represented by physical descriptors such as particle size and aspect ratio or statistical descriptions based on the two-point correlation function. Alternatively, it was characterized by features extracted via the convolutional layers in a CNN model where different filters are applied to obtain different features.

The other important part is microstructure reconstruction, which is the reverse process of microstructure characterization. It enables the generation of 2D and 3D microstructures with prescribed microstructural features (i.e., those based on the above introduced three types of descriptors) \cite{bargmann2018generation,reconstruction2006,seibert2022microstructure}. Depending on the microstructure characteristics, microstructure generation could be implemented by various algorithms \cite{bargmann2018generation,seibert2022microstructure}.
For instance, Seibert et al.~\cite{seibert2022microstructure} introduced a powerful platform for microstructure characterization and reconstruction, i.e., MCRpy. The workflow of MCRpy is depicted in Fig~\ref{fig:microstructure reconstruction}b. It treats the microstructure reconstruction as a modular and adaptable optimization problem. And it enables the consideration of diverse descriptors for microstructure characterization, arbitrary loss functions, and various optimizers for microstructure reconstruction. Recently, it is demonstrated that ML models such as the generative adversarial network (GAN) models can efficiently reconstruct complicated microstructures (e.g., \cite{xu2021guiding,zhang2024vegan}).
More information on microstructure characterization and reconstruction is found in a relevant comprehensive review paper \cite{bostanabad2018computational}.

\subsection{Fundamentals of Machine Learning}
Before introducing how ML methods can advance the design and modeling of microstructured materials, it is necessary to understand its fundamentals and how it operates in practice.
ML is a structured process that can be outlined by a general workflow regardless of tasks and model architectures. As illustrated in Fig.\,\ref{fig:machine learning}b, the general workflow includes:
\begin{enumerate}[I.]
    \item Task definition: A ML process begins with defining the task and the objective function. Depending on the tasks and objectives, ML can be broadly categorized into three learning paradigms: supervised learning, unsupervised learning, and reinforcement learning. The different tasks can be broadly defined as regression, classification, dimensionality reduction and recommendation, as described in many standard ML lecture notes and books (e.g., \cite{Goodfellow-et-al-2016}). While other types of tasks are equally important, we focus on regression problems as they are closely related to material modeling and design of microstructured materials in this paper. 
    \item Data strategy plan: Data collection steps follow, and preprocessing is then performed to handle missing values, outliers, and inconsistencies. Data scaling is necessary if the inputs and outputs vary over magnitudes. The data is usually split into training and test sets.
    \item Model selection: A suitable ML model is selected or developed based on the task and data. Models can be generally classified into two main types: conventional models (shallow learning models) and deep learning models. Conventional models further fall into parametric and non-parametric categories. Parametric models have a defined number of parameters, while non-parametric models experience an increase in the number of parameters with expanded input data~\cite{murphy2012}. One of the most simple ML models is the linear model. Although it is brief, the linear model already includes the main features of a ML model. It serves as a mathematical model describing the overall linear relationship between a set of inputs and the corresponding outputs. In this context, model training involves determining the parameters (i.e., the slope and intercept) in the linear model to best fit the given data points. Essentially, the goal is to minimize the difference between the model prediction and the reference outputs for the given inputs. Other typical conventional ML models include Decision trees ~\cite{DT}, Random Forest, Support Vector Machines, K-Nearest Neighbors, Gradient Boosting models, and Gaussian Processes (see e.g., \cite{wang2022data}). 
Different types of neural network (NN) models are typical deep learning models. The architecture of NN models usually consists of one input layer connected to an output layer by a number of connected hidden
layers. Each layer possesses a number of neurons. Each neuron is characterized by a weight, a bias, and possibly an activation
function (e.g., sigmoid, tanh, and relu) to introduce non-linearity. The depth and width of the layers are critical factors. Depth refers to the number of hidden layers, while width relates to the number of neurons in each layer. The trade-off between depth and width is crucial, as a large width can hinder the learning capability of the model. Achieving an optimal balance between the two is key for efficient and accurate prediction performance.

Due to their powerful fitting capability, NN models are widely used for data-driven material modeling and design. Artificial neural network models (ANNs) are commonly used when the inputs are explicitly defined, e.g., microstructures represented by well-defined geometrical parameters or statistical features. 
Convolutional Neural Networks (CNNs), Recurrent Neural Networks (RNNs), Generative Adversarial Networks (GANs), and Graph Neural Networks (GNNs) are among other well-known types of neural networks. CNN models are suitable for situations where microstructure images act as the inputs. RNN models are capable of modeling history-dependent physical problems, e.g., viscoelastic and plastic problems. GAN models can be used to generate new microstructures with features similar to those in the training dataset. GNN models are preferred when the microstructures can be converted to a graph consisting of nodes connected by edges, such as polycrystalline microstructure and lattice structures. Besides, deep language models have also been exploited to address different problems relevant to materials design and modeling~\cite{Goodfellow-et-al-2016,hu2023deep}.
    \item Training:    
    The fundamental function of the ML workflow is its ability to learn, facilitated by a relatively straightforward recipe and the utilization of two key ingredients: data and a model. Additionally, an objective function, such as mean squared error or cross entropy, is employed. This function, also known as an error, cost, or loss function, calculates the average loss across the entire training dataset. In essence, training involves an iterative optimization process aimed at enhancing the model's learning capabilities. It should be noted that training is more than a pure optimization in the mathematical sense. In traditional optimization, the objective is to minimize/maximize a specific function based on the available data. However, in ML, the objective is to minimize the error or loss of unseen data, namely the generalization error. Hence, ML involves specific generalization strategies, which are intentionally designed to sacrifice training error for generalization error. For instance, methods are employed to modify the cost function, by introducing a penalty term (L1 or L2 type) into the loss function~\cite{Goodfellow-et-al-2016}. To efficiently train a ML model with a satisfactory performance, hyperparameters such as the learning rate and the batch size should be properly chosen.  Note the difference between hyperparameters and parameters. Hyperparameters are prior choices for the specification of model architecture (e.g., layer and neuron numbers of a NN model) or for defining training procedures (e.g., learning rate and regularization-related ones), while the parameters of a model are determined through training.  
    \item Performance evaluation: Once a ML model is trained, its performance is evaluated using predefined evaluation metrics on the validation or test data set. This performance evaluation helps to understand whether the model is underfitting, overfitting, or performing optimally. It guides the hyperparameters tuning to attain a model with satisfactory performance. 
    \item Deployment: Once the model is ready, it undergoes packaging, versioning, and integration into
the production environment. Continuous monitoring of the model’s performance with incoming data is essential, and necessary updates
are made based on new data or requirements. \\
\end{enumerate}

In practice, the best type of ML models and the corresponding hyperparameters to choose depend on the characteristics of the tasks to solve, the nature of the data (e.g., amount and distribution), and the available computational sources. A good practice is to test multiple models and select the one with the best performance. Experienced practitioners usually have good sense of choosing hyperparameters, selecting/comparing various models along with flexible data strategies. 

\section{Data-driven Multiscale Simulation}
\label{Mult_Simu}
The literature review underscores the substantial potential of ML to enhance materials modeling, highlighting a critical gap in microstructure-informed materials modeling. While conventional constitutive models for homogeneous materials are notably advanced and closely aligned with experimental observations, ML-based material models in the literature often serve as alternatives or complements to these established models. However, the development of microstructure-informed models for microstructured materials remains a significant challenge, with few conventional models available. In this scenario, the complexity and diversity inherent in microstructures, along with their complicated material behavior, surpass the capabilities of conventional methods but offer an ideal platform for ML-based modeling approaches. As summarized in Table~\ref{tab1}, for both the cases of homogeneous materials and microstructured materials, the ML-based material model takes specified physical variables (e.g., physical fields, driving forces, or tensor invariant) as inputs and predicts the corresponding physical outputs (e.g., thermodynamically conjugated quantities, and energy/thermodynamical potentials) or effective material properties. Here we focus on the latter case of microstructured materials for which microstructural descriptors may also be introduced as inputs. In the subsequent discussion, we provide a summary of current research initiatives aimed at addressing how these ML-based material models are integrated into multiscale frameworks for modeling and simulation of microstructured materials. In general, existing works on data-driven multiscale simulation are classified as sequential and concurrent types. In the sequential case, the data-driven method is adopted to predict the effective (macroscopic) material properties or construct surrogate effective material models. In the concurrent case, the constructed data-driven surrogate material model is further incorporated into the macroscale simulations. In the following, we will introduce some typical examples for each case.

\begin{figure}
\centering
  \includegraphics[width=0.78\columnwidth]{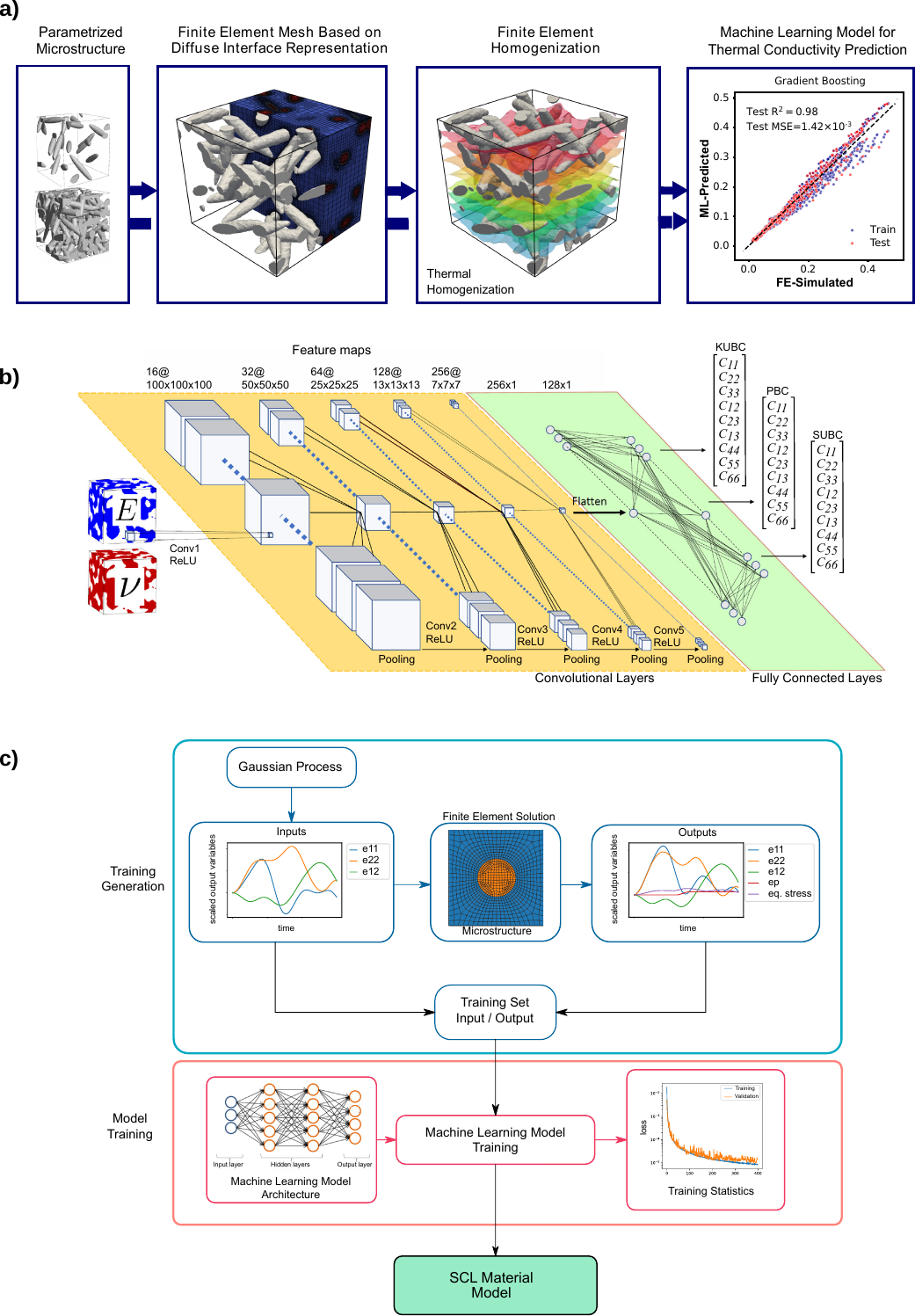}
 \caption{ML-aided sequential multiscale simulation. a) Predicting the effective thermal conductivity of composite structures as a function of microstructural parameters and base material properties~\cite{fathidoost2023data}. b) CNN-based modeling of microstructure-dependent effective stiffness tensor of 3D heterogeneous microstructures~\cite{eidel2023deep}. c) Surrogate modeling of elasto-plastic constitutive relations in microstructured materials~\cite{LOGARZO2021}. Figure a) is adapted from~\cite{fathidoost2023data} (own work published under license CC BY-NC-ND 4.0 \url{https://creativecommons.org/licenses/by-nc-nd/4.0/}). Figure b) is adapted from~\cite{eidel2023deep} (published under license CC BY 4.0 \url{https://creativecommons.org/licenses/by/4.0.}). Figure c) is adapted from~\cite{LOGARZO2021} with permission from Elsevier.}
\label{fig:sequential multiscale}
\end{figure}

\subsection{Sequential Scheme: Effective Property Prediction}
ML methods have been widely applied to model the microstructure-dependent effective properties in microstructured materials. In these cases, the form of the effective material model is usually known in prior and the corresponding unknowns are the effective material parameters. For instance, effective linear elastic behavior is characterized by the effective Hooke's law, and the effective thermal conducting is described by the effective Fourier's law. In this regard, the ML model takes the microstructure features and/or base material properties as inputs and predicts the corresponding effective material properties. 

The authors have contributed several works in this field.
Fathidoost et al.~\cite{fathidoost2023data} developed a gradient boosting-based ML surrogate model to predict the effective thermal conductivity of particulate composite structures. The microstructure features including the aspect ratio and volume fraction of particles and the base material thermal properties including the interface thermal resistance and thermal conductivity of the base materials are considered as the inputs (see Fig. \ref{fig:sequential multiscale}a). In Peng and Xu~\cite{peng2024data}, ANN models are constructed to predict the effective linear elastic properties including the effective Young's and shear moduli and Poisson's ratio of composite triangular lattice structures, where the structural features including the strut thickness, strut angle, and Young's modulus ratio between the two base materials.

In the above two examples, the microstructure is well-defined by a few microstructural descriptors. For disordered microstructures, more advanced ML models should be exploited. 
As shown in Fig.\,\ref{fig:sequential multiscale}b, a CNN model is proposed to predict the effective stiffness of 3D random heterogeneous multi-phase materials, where the microstructure images act as the inputs. Peng and Xu~\cite{peng2024poly} considered the data-driven surrogate modeling of effective ionic conductivity of polycrystalline oxide ceramics with varying microstructural features. An efficient GNN model is exploited to capture the features of individual grains and grain boundaries as well as their interactions. There are also other works on applying ML method to identify the effective properties of microstructured materials, e.g., effective elastic properties of multi-material metamaterials~\cite{kulagin2020architectured,pahlavani2022deep}, effective Young's modulus of metal foams~\cite{chen2022ai}, and effective magnetostriction of polycrystalline materials \cite{dai2021graph}. 

These surrogate models can act as the efficient property predictor for the optimization and inverse design of microstructured materials (see Section~\ref{optimization_inverse}). However, the methodology discussed here only applies to the cases where apparent macroscopic material parameters are of concern.

\subsection{Sequential Scheme: Effective Constitutive Surrogates}
The effective material behavior of microstructured materials is usually dramatically different from the underlying base materials and, hence can hardly be characterized by already known material models. In this regard, ML models can act as surrogate effective material models. As summarized in Table~\ref{tab1}, in this scenario, the inputs of the ML model are physical variables such as physical fields, driving forces or tensor invariant and/or microstructural features, while the outputs are thermodynamically conjugated quantities or energy/thermodynamical potentials. In Fig.\,\ref{fig:sequential multiscale}c, one example~\cite{LOGARZO2021} is shown to illustrate the general workflow to construct a data-driven surrogate material model. In this work, the effective plastic behavior of a composite is considered. Numerous finite element (FE) simulations at the RVE level are performed to collect the dataset consisting of strain sequences paired with the corresponding stress sequences. Then, a RNN model with time-history prediction capability is constructed and trained with the dataset. The trained RNN model is essentially an incremental elastic-plastic constitutive model. Fernández et al. \cite{fernandez2022material} developed an ANN-based hyperelastic constitutive model for cubic lattice structures undergoing large deformation where both the structure parameters and the strain tensor act as the inputs and the output is the elastic potential. This ML-based material model can capture the strongly nonlinear deformation behavior of the lattice structures with varying structure parameters. Other related works are found in the review paper on data-driven multiscale modeling~\cite{bishara2023} and the references therein. So far, only very limited works considered the microstructure dependence in the ML model. One reason is that much more data points are required to train the ML model with the microstructure descriptors as the additional inputs.

\subsection{Surrogates-enabled Concurrent Scheme}\begin{figure}
\centering
  \includegraphics[width=0.85\columnwidth]{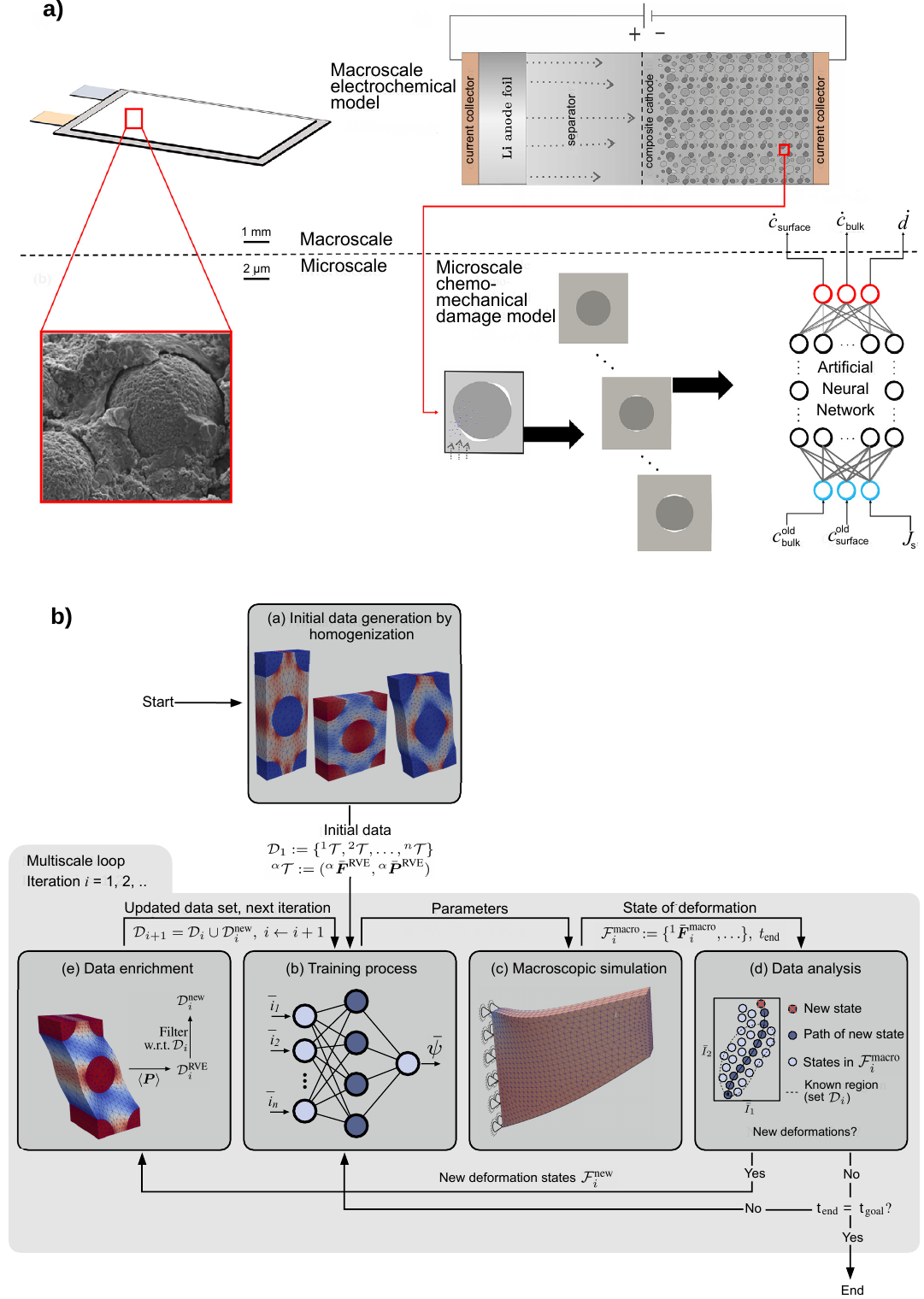}
 \caption{ML-aided concurrent multiscale simulation. a) Two-scale damage analysis in Li-ion batteries~\cite{asheri2023}. b) Finite strain hyperelastic behavior of composites \cite{kalina2023fe}. Figure a) is adapted from~\cite{asheri2023} with permission from Elsevier. Figure b) is adapted from~\cite{kalina2023fe} (published under license CC BY 4.0 \url{https://creativecommons.org/licenses/by/4.0.}). }
\label{fig:concurrent multiscale}
\end{figure}

Here, the data-driven concurrent scheme refers to the cases where the trained surrogate models are further incorporated into the simulations at the macroscale. Two such examples are shown in Fig.\,\ref{fig:concurrent multiscale}. 
As displayed in Fig.\,\ref{fig:concurrent multiscale}a, Asheri et al.~\cite{asheri2023} proposed a data-driven multiscale strategy to predict battery cell performance to bypass the high computational cost encountered by the conventional FE$^2$ method. A large dataset is obtained from simulations at the microscale for electrolyte-active material two-phase RVEs with varying material and state variables. The influence of the interface damage between the solid electrolyte and cathode active material which plays a critical role in cell performance and degradation is considered in a coupled chemo-mechanical model. Based on the dataset, an ANN-based surrogate model is constructed and trained, which can efficiently predict microscopic chemo-mechanical behavior.
The surrogate model is further integrated into a multi-field two-level framework to predict cell performance, focusing on the impact of interface damage on capacity loss. The performance of this data-driven multiscale strategy is compared with the results from the original multi-field two-level simulations. The comparison demonstrates that the data-driven multiscale approach shows promising results with high accuracy and efficiency, highlighting the potential of data-driven multiscale simulations for lithium-ion battery design. 

Kalina et al.~\cite{kalina2023fe} (Fig.\,\ref{fig:concurrent multiscale}b) developed a data-driven multiscale approach based on physics-constrained NN models and automated data mining. To showcase the methodology, they consider the effective hyperelastic behavior of a fiber-reinforced composite. The physics-constrained ML model takes a set of deformation tensor invariants as inputs and outputs the corresponding free energy density. Unlike most data-driven multiscale methods, the ML model will be re-trained during the simulation process if newly unseen deformation states are observed. This is more efficient than directly training the ML model with all the possible deformation states expected to occur in the macroscopic deformation. The multiscale loop includes model training, macroscopic simulation, data analysis and data enrichment. 

Mianroodi et. al.~\cite{mianroodi2022lossless} considered the ML-aided multiscale modeling of nanoporous materials. To this end, a CNN model with microstructure images as inputs and the corresponding elastic properties as outputs is trained based on data attained from molecular statics calculations. This CNN model is further incorporated into upper-scale FE simulations as the constitutive law. This ML-based multiscale simulation is much more efficient than the full atomistic simulation.

More related works are found in the review paper on data-driven multiscale modeling~\cite{bishara2023} and the references therein. Multiscale modeling of composites with spatially varying microstructures deserves more effort. In this regard, the microstructure descriptor should be considered as additional inputs in the surrogate model.

In general, the data required to train the ML-based material models can be collected from experimental tests or generated by numerical simulations with specified microscopic physical models. On the one hand, generating data from simulations is usually much cheaper and enables high accuracy by precisely controlling the relevant conditions and parameters. On the other hand, experimental data that does not rely on assumptions and physical models can better reflect real-world material behavior, and hence the resultant trained ML models can be generalized and deployed for real-world applications. Due to the limited available experimental data, in most existing works, only simulation data is employed to train ML models. Although it may not be feasible to solely rely on experimental data, a combination of simulation and experimental data along with appropriate data augmentation techniques is necessary to attain generalizable ML models for real-world applications.

\section{ML-enabled Microstructure Correlation and Sensitivity Analysis}
\label{Correlation_sensitivity}
In the process of fabrication or synthesis of microstructured materials, various process parameters are involved. These parameters determine the resulting microstructures. The latter further affects the macroscopic properties of the produced materials. Thus, revealing the intricate relations among processes, microstructures, and properties enables the design and fabrication of microstructured materials with target properties~\cite{baskaran_micros2021,CHUNG_sps2017}. These relationships are usually strongly nonlinear with various inputs and outputs, and hence, are challenging for traditional methods to tackle. In contrast, ML models with powerful fitting capability are well-suited for these tasks.

The ML methods can be applied to construct process-microstructure (PM), microstructure-property (MP), process-microstructure-property (PMP), and process-property (PP) correlations. 
In this paper, we focus on MP and PM correlations. 
Although a large number of process/microstructure parameters are involved in these correlations, they are not equally important. In this regard, the trained ML surrogate models can be exploited for sensitivity analysis, which enables the identification of most relevant parameters for the corresponding correlations~\cite{SA_thermal,lin2021data}. In the following, we will address these aspects by introducing typical examples.
\subsection{Microstructure-Property Relation}
Here we will introduce how ML methods enable surrogate modeling of MP relation by typical examples. Cases with different types of microstructure characterization are separately discussed (see Section~\ref{sec:Microstructure Characterization}).
\begin{figure}
\centering
  \includegraphics[width=0.85\columnwidth]{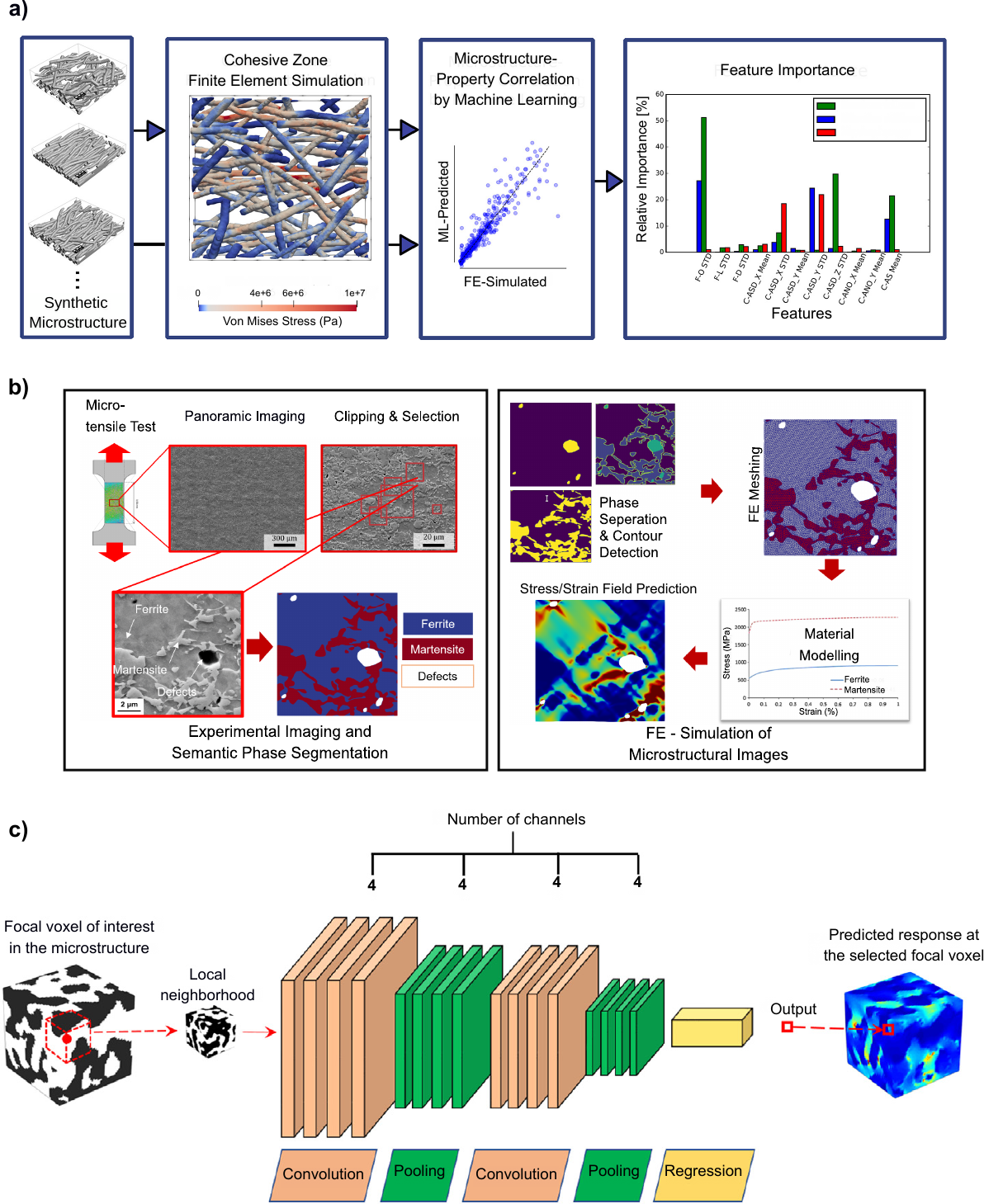}
 \caption{ML-aided modeling of microstructure-property relation. a) Microstructure-dependent mechanical properties of paper materials~\cite{lin2021data}. b) CNN-based end-to-end prediction of the mechanical field in 2D microstructures of dual-phase steel under tensile loading~\cite{lin2023machine}. c) CNN-based prediction of microscale elastic strain field of 3D voxel-based microstructure of a high-contrast two-phase composite~\cite{yang2019establishing}. Figure a) is adapted from~\cite{lin2021data} (own work published under license CC BY-NC-ND 4.0 \url{https://creativecommons.org/licenses/by-nc-nd/4.0/}). Figure b) is adapted from~\cite{lin2023machine} (own work published under license CC BY 4.0 \url{https://creativecommons.org/licenses/by/4.0.}). Figure c) is adapted from~\cite{yang2019establishing} with permission from Elsevier.}
\label{fig:MPR}
\end{figure}
\subsubsection{Descriptor-based}
In this scenario, a microstructure is characterized by a few physical descriptors. Xu et al.~\cite{xu2014descriptor} considered MP relations of two-phase polymer composites. The microstructures are represented by the volume fraction of inclusion, the average nearest cluster center distances, and the distribution of equivalent radius and elongation ratio of the inclusion, which quantify the composition, dispersion status, and geometry, respectively. A data-driven surrogate model is constructed to correlate the viscoelastic properties of the composites with these microstructural features. 

As reviewed in~\cite{xu2021guiding}, the descriptor-based approach is widely used to characterize the Li-ion battery microstructures. In this scenario, various microstructural features are involved. The composition category includes the volume fraction of active materials, binders, additives, and pores. The specific surface and interface areas, surface-to-volume ratio, connectivity, and tortuosity belong to the dispersion/agglomeration category. Particle size (volume, equivalent diameter, diameter/radius along different axes), type of particle shape/morphology/profile (sphere, ellipsoid, cylindrical, etc.), sphericity, particle orientation, surface roughness, and pore size/radius/diameter are among the geometry category. The correlation between the microstructural features and the electrochemical and mechanical properties of Lithium-ion battery materials are captured by ML models. 

One of the works from our group~\cite{lin2021data} employed the feature-based ML approach to correlate the paper microstructures with their mechanical properties (Fig.\,\ref{fig:MPR}a). To this end, numerous microstructures with varying features are generated and analyzed. Several features are extracted. FE simulations are then conducted to determine their mechanical properties including the failure strain, effective stiffness, and maximal stress under tensile testing. The resulting dataset consisting of mechanical properties paired with the structural features is adopted to train a ML model. This ML model is exploited to conduct the sensitivity analysis to evaluate the importance of different features on each mechanical property.
\subsubsection{Correlation-function-based}
The n-point correlation functions are commonly adopted to statistically characterize disordered heterogeneous materials. In~\cite{cheng2022data}, a data-driven framework is proposed to identify n-point correlation functions for characterizing a number of microstructured materials including Sandstone, metal-ceramic composites, Pb-Sn alloys, concrete, and particle-reinforced composites. It is shown that the n-point correlations can be computed by image convolution. Thus, a CNN model is constructed. The microstructure reconstruction is considered as an optimization problem, i.e., finding a microstructure with correlations matching the target correlations. A searching algorithm is introduced to select a concise subset of n-point correlations. The optimized microstructure representations are directly used to compute material properties such as bulk/shear modulus and thermal/electrical conductivity by adopting the effective medium theory. In such a manner, MP relations can be built with significantly less data in comparison with purely data-driven methods.

In~\cite{hao2024novel}, 2-point correlation functions have been exploited to characterize microstructures of SiC fiber-reinforced titanium matrix composites. To this end, three correlation functions including two for matrix/fiber auto-correlations and one for matrix-fiber cross-correlation are introduced. PCA is further adopted for dimensionality reduction so as to attain reliable low-dimensional descriptors. FE simulations are conducted to obtain the ultimate tensile strength of the microstructures, which together with the microstructural features and the base material properties comprise the dataset. With the aid of this dataset, a support vector regression ML model is constructed and trained for the relevant MP correlation. The ML model is further validated by experimental data.
\subsubsection{ML Representation-based}
Instead of relying on statistical or physical microstructure features, the microstructure image itself can serve as the input for establishing the MP relation. In this regard, all the microstructure information is resolved. CNN models are commonly adopted for image-based MP correlation in 2D and 3D microstructured materials. 

A 2D example is illustrated in Fig.~\ref{fig:MPR}b. In this work by Lin et al.~\cite{lin2023machine}, CNN-based surrogate models are constructed to predict von Mises stress and equivalent plastic strain fields in commonly used dual-phase steels for automotive applications. The ML model operates in an end-to-end manner, taking 2D segmented phase images from actual experimental scanning electron micrographs as inputs and providing the mechanical fields as outputs. The CNN models are trained using the U-net NN structure with data obtained from numerous elastoplastic FE simulations of various microstructure samples. The model performs well even for microstructures unseen in the training dataset. A 3D example~\cite{yang2019establishing} is shown in Fig.~\ref{fig:MPR}. In this work, a CNN model is applied to predict the elastic strain field in 3D high-contrast two-phase microstructure. It is demonstrated that CNN-based ML method is a promising technique to construct feature-engineering-free models for MP correlations in sophisticated microstructured materials. These two works on adopting CNN model for prediction of the mechanical field distribution belong to category III in Table~\ref{tab1}.

One drawback of CNN-based surrogate models for MP relation is the high computational cost due to the high dimensionality of the image-based inputs. This issue is more evident for 3D microstructures.
\subsection{Process-Microstructure Relation}
\begin{figure}[p]
    \centering
\includegraphics[width=0.85\columnwidth]{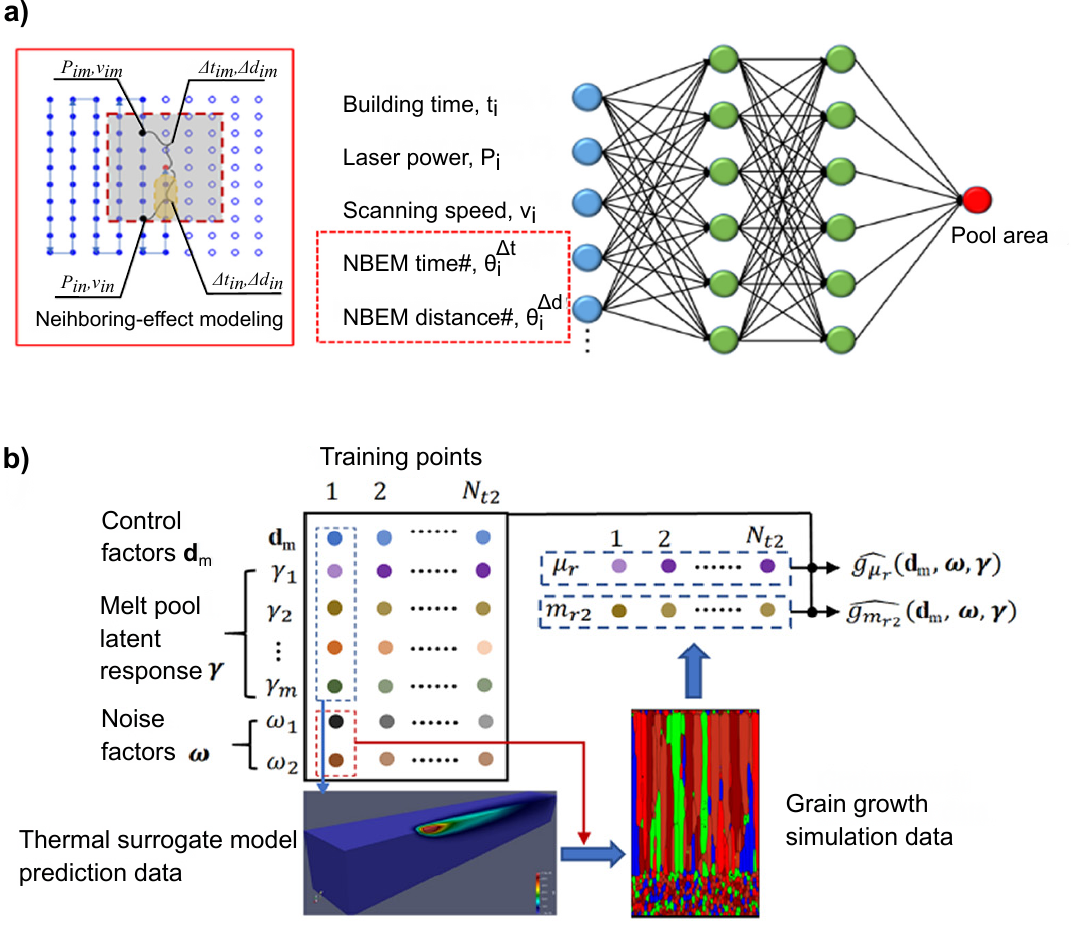}
    \caption{Data-driven modeling of process-microstructure relation. a) Data-driven modeling of correlation between process parameters and melting pool features~\cite{wang2022data}. b) Data-driven modeling of grain structures as functions of process parameters~\cite{wang2022data}.
    Figures a) and b) are adapted from ~\cite{wang2022data} with permission from Elsevier.}
    \label{process}
\end{figure}
Fabricating or synthesizing materials with sophisticated microstructures is the final step before their practical applications. These tasks are highly challenging by conventional methods, especially when it comes to microstructures with controllable features. In recent years, various additive manufacturing (AM) techniques are rapidly developing. These ground-breaking technologies enable the fabrication of complex microstructured materials with different types of base materials. An AM process is usually governed by a number of process parameters, which strongly affect the characteristics of the produced microstructures and thereby their properties. Thus, it is vital to understand the intricate process-microstructure correlations relevant to AM. In this regard, ML methods can play important roles. There are already a few review papers~\cite{wang2022data,debroy2021metallurgy} which well summarize some recent progresses in this field. A few representative examples are introduced here.

The first step before starting an AM process is selecting the process parameters such as the feeding rate, scanning speed, laser power and laser parameters for producing a part with target features~\cite{wang2022data}. The intermediate quantities such as the pool properties, the bead properties, and the temperature fields and history can be directly correlated to these process parameters by data-driven surrogate modeling (see Fig.\,\ref{process}a)~\cite{xiong2014bead,kamath2016data,yang2020scan,wang2020uncertainty}. 

Monitoring and controlling of manufacturing process is of great importance to perform on-the-fly quality control of the produced parts. Imani et al.~\cite{imani2019deep} develop a deep NN model to identify incipient flaws from layer-wise spatial images. The good performance of the model on real-time flaw detection is validated by experiments. This presents an opportunity to identify initial defects in AM processes before completion.

Process parameters can also be directly correlated to the final microstructural features, such as grain structures~\cite{wang2019data} (Fig.\,\ref{process}b), porosity~\cite{xie2022data}, and geometrical distortion~\cite{zhu2018machine,hong2021artificial}. 

In most existing works on ML-enabled microstructure correlation, data is generated by different types of simulations. Incorporating experimental data into the ML model training is highly desired to enhance their performance in practical applications.
\section{Microstructure Optimization and Inverse Design}
\label{optimization_inverse}
Optimization and inverse design of microstructures are two different conceptions often mixed up in the literature. The former is aimed at finding a microstructure with extreme properties (i.e., minimum or maximum) under certain constraints. However, the latter focuses on attaining a microstructure with specified target properties which are not necessary to be extreme. 

Both tasks pose significant challenges due to their inherent high dimensionality, strong nonlinearity, and computational demands. However, ML methods offer efficient solutions to these challenges. In this section, we will illustrate by representative examples how to exploit ML methods to realize microstructure optimization and inverse design.
\subsection{Microstructure Optimization}
\begin{figure}[p]
    \centering
\includegraphics[width=1\textwidth]{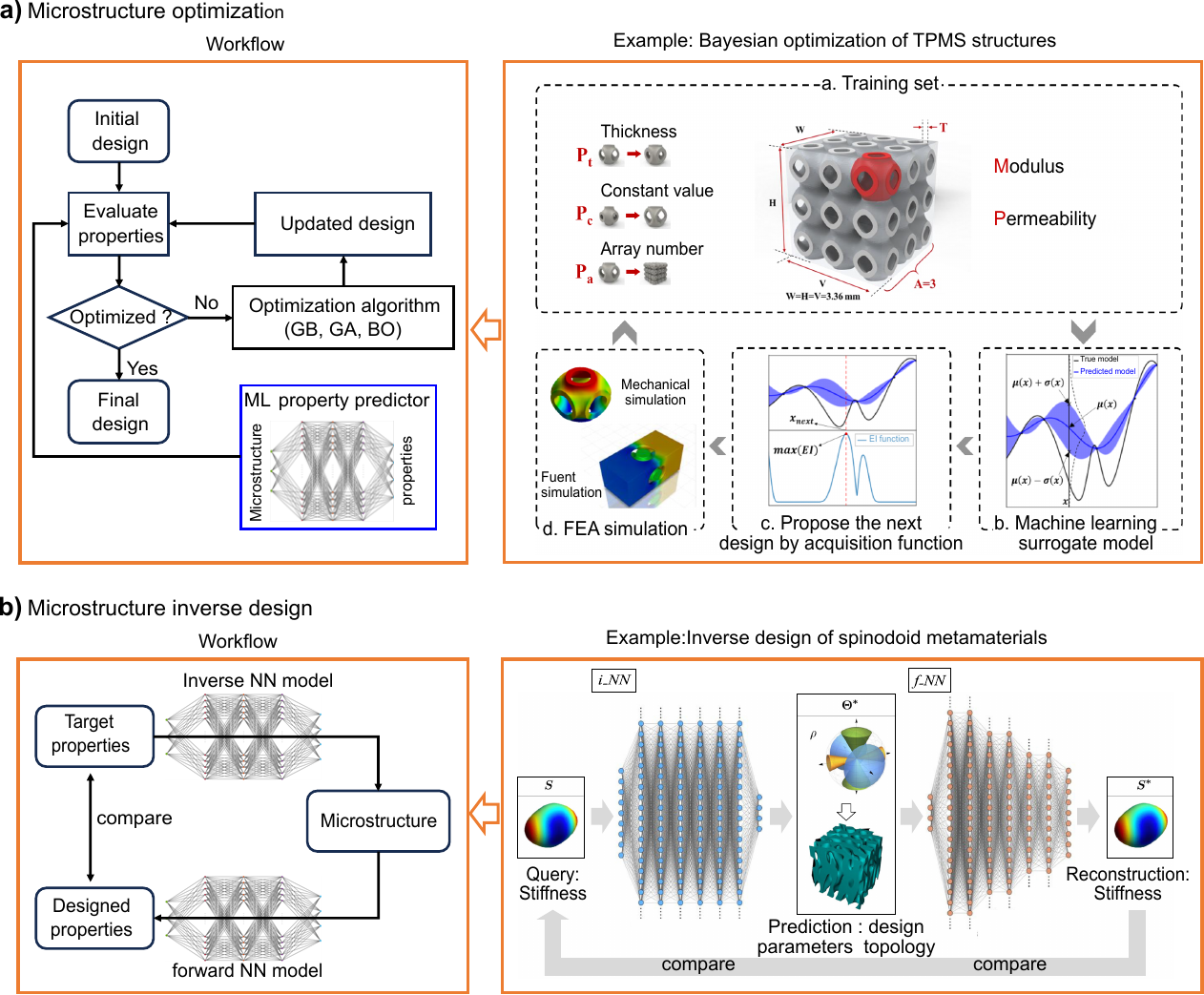}
    \caption{Data-driven microstructure a) optimization and b) inverse design: Workflows and typical examples (\cite{hu2023multi,kumar2020inverse}).
   The right part of Figure a) is adapted from~\cite{hu2023multi} with permission from Elsevier. The right part of Figure b) is adapted from \cite{kumar2020inverse} (published under license CC BY 4.0 \url{https://creativecommons.org/licenses/by/4.0.}).}
    \label{inverse}
\end{figure}

Optimizing microstructures to achieve extreme effective properties is a common demand in the design process. As illustrated in Fig.~\ref{inverse}a, the optimization is generally an iterative process including proposing initial designs and then updating them until some specified convergence criteria are fulfilled. Depending on the applied algorithms, optimization methods can be categorized as gradient-based (i.e., local) or gradient-free (i.e., global) types. In both cases, the time-consuming property predicting is involved in each iteration. In ML-based methods, property predicting is achieved by a trained ML model, which speeds up the optimization process. In the following, we review different ML-based optimization methods.
\subsubsection{Gradient Based-local Method}
The topology optimization method is a frequently used gradient-based optimization method to find out an optimized structure by tuning the material or microstructure distribution. As a grade-based method, at each iteration, the sensitivity of the objective function with respect to the design parameters must be evaluated, which is usually achieved by solving certain boundary value problems in a finely meshed domain. This makes the method computationally expensive. Furthermore, in the multiscale topology optimization case, the situation is even worse. At each iteration, similar to the multiscale problem discussed in Section~\ref{Mult_Simu}, the effective properties of the microscopic properties are required at each material point in the macroscopic simulation. As already illustrated by some recent works, data-driven methods can help to remedy these issues and thereby dramatically increase computational efficiency. Senhora et al.~\cite{senhora2022machine} proposed an efficient ML-based topology optimization framework, where a CNN-based surrogate model is used to predict the sensitivity for a given intermediate design, which speeds up the optimization process. Recently, the data-driven multiscale topology optimization method has been successfully exploited to design optimized macroscopic structures based on different types of microstructures such as isotruss metamaterials~\cite{white2019multiscale}, spinodoid metamaterials~\cite{zheng2021data}, and strut-based lattice structures~\cite{wang2022data}. In these works, ML models are introduced to map the microscopic design parameters to the effective properties at the macroscopic material point. Furthermore, the differentiable ML models can efficiently provide the sensitivity of the effective properties with respect to the design parameters, which further reduces the computational cost.

\subsubsection{Global Method: Bayesian Optimization}
Bayesian optimization is a data-driven method to optimize a black-box objective function with a limited number of function evaluations. It consists of two main parts: a statistical surrogate model (e.g., Gaussian Process) for the objective function and an acquisition function based on the former for determining the next sampling point towards maximizing the objective function. The statistical surrogate model initially trained by a few data points is re-trained after adding each newly sampled data point. The acquisition function results from the trade-off between the exploitation and exploration. After a number of iterations, even with a few data points, Bayesian optimization can already help to find out a design with improved properties. Thus this method is suitable for the cases where only a limited number of data points are available. The example shown in Fig.\,\ref{inverse}a~\cite{hu2023multi} illustrates the general workflow of Bayesian optimization. The optimization of the strength and permeability of triply periodic minimal surface (TPMS) structures is considered. To this end, the structural parameters and properties of a few initial TPMS structures are evaluated by simulations to generate the initial dataset, which is subsequently applied to train the ML surrogate model. Then, the recommended design is proposed based on the acquisition function. The properties of the new design are calculated by simulations. The updated dataset with the newly generated data is used to re-train the ML model. The above steps are repeated until the optimized structure is found. 

Bayesian optimization has also been widely used to optimize properties of  other microstructured materials, e.g., the transport properties of electrode materials~\cite{duquesnoy2023machine}, the plastic properties of dual-phase steel microstructures~\cite{kusampudi2023inverse}, the mechanical properties of hexagonal honeycomb metamaterials~\cite{kuszczak2023bayesian}, the rotation thermal modulation contrast ratio of multilayer hyperbolic metamaterials~\cite{liao2024data}, and the transport properties of graphene thermoelectrics~\cite{yamawaki2018multifunctional}.

Bayesian optimization also has some limitations, e.g., it becomes extremely computationally expensive when dealing with problems with high-dimensional design spaces. 

\subsubsection{Global Method: Genetic Algorithm}
The genetic algorithm is a typical gradient-free optimization method. It starts with an initial population consisting of a number of randomly generated designs. Then, the initial population is subjected to selection, crossover and mutation to generate the updated population. This process is repeated until the optimized design is attained. During the whole process, the evaluation of the properties of the new designs are continuously desired, which could be time-consuming. In this regard, the ML surrogate model can serve as the efficient property predictor once been trained. Recently, by integrating ML method with the genetic algorithm, the optimization of different effective properties of microstructured materials are realized, e.g., the buckling resistance of truss lattice materials~\cite{maurizi2022inverse}, the customized loading curves of the shell-based mechanical metamaterials~\cite{wang2022inverse}, the effective Poisson’s ratio of perforated auxetic metamaterials~\cite{liu2023high}, and the fracture toughness of polycrystalline materials~\cite{hsu2021tuning}. 

\subsection{Microstructure Inverse Design}
Instead of finding a microstructure design with extreme properties, the inverse design focuses on predicting a design with specified target properties. In general, the data-driven optimization method can also be applied for inverse design purposes provided the objective function is suitably selected. However, by the optimization-based inverse design method, an iteration process is required for each inverse design task. A more efficient alternative is to construct a surrogate model which takes the target properties as inputs and predict the corresponding microstructure design as outputs. However, due to the fact that multiple microstructures may correspond to the same effective properties, the inverse design task is generally ill-posed. This issue can be remedied by the data-driven inverse design method. Liu et al.~\cite{liu2018training} proposed a tandem network model by combining a forward and an inverse neural network models. The workflow of this data-driven inverse design method is illustrated in Fig.~\ref{inverse}b by a typical example~\cite{kumar2020inverse} where the inverse design of effective elasticity of spinodoid metamaterials is investigated. The forward NN model acting as the property predictor is trained first. Then, the inverse NN model is trained by minimizing the reconstruction loss (i.e., the error between the properties of the designed structure and the target values) instead of the prediction loss (i.e., the error between the predicted microstructural features and those corresponding to the inputs), where the reconstruction loss is evaluated by the already trained forward NN model. This ML-based direct inverse design strategy has been successfully applied to realize the inverse design of different materials such as the transmission spectrum of nanophotonic structures~\cite{liu2018training}, the deformed configuration of shape-programmable 3D kirigami metamaterials~\cite{alderete2022machine}, the effective elasticity of truss metamaterials~\cite{bastek2022inverting}, the effective elasticity of composite triangular lattice structures~\cite{peng2024data}, and the auxeticitiy of curved beam metamaterials~\cite{felsch2023controlling}.

The introduced data-driven inverse design method based on ANN models usually requires a large dataset to train models with sufficiently high accuracy. Currently, in most cases, the dataset is generated by inexpensive numerical simulations. This method may be less efficient when the available data is limited.

In summary, data-driven optimization/inverse design methods can efficiently and accurately solve the optimization/inverse design tasks for various microstructured materials with respect to different properties. Each method has its own advantages and limitations. The reviewed representative works for different methods can provide useful instructions on how to select the most suitable one for a specified problem.

\section{Concluding Remarks and Perspectives}
\label{concluding}
In this paper, the recent contributions on applying ML methods to advance the modeling and design of microstructured materials are reviewed. It is illustrated that the ML methods can play their important roles in various aspects relevant to the microstructure-informed materials modeling including microstructure characterization and reconstruction, multiscale simulation, process and/or microstructure-property correlation, 
 and microstructure optimization and inverse design. Although significant progress has been achieved in this field, there are still important aspects desiring further exploration. Based on our knowledge and understanding, we present the following perspectives:
\begin{enumerate}[a)]
     \item In the data-driven multiscale simulation framework, most ML-based surrogate models  are only applicable for a specified microstructure. To construct a surrogate model generally applicable to at least a class of microstructures, microstructural features should be considered as inputs in the ML models. In this regard, a much larger dataset should be generated by sampling both the loading conditions and the microstructural features. To reduce the number of required data, one could adopt the automated data-mining strategy as suggested in \cite{kalina2023fe}. 
    \item In the context of process-microstructure-property correlation, most works focus on data-driven surrogate modeling of MP correlations. Limited works consider the PM correlation. The direct PP correlation is rarely explored. Future endeavors can be devoted to bridging the gaps and exploring the direct linkage between the manufacturing process and resulting material properties.
    \item In the context of microstructure optimization and inverse design, most works focus on microstructures that can be well-defined by a limited number of features. Addressing the more intricate task of optimizing and inversely designing sophisticated microstructures demands more effort. Accomplishing this challenging task requires the incorporation of more advanced ML models. A noteworthy recent endeavor in this direction is presented in \cite{kusampudi2023inverse}, where a generative ML model is combined with Bayesian optimization to achieve the inverse design of polycrystalline dual-phase steels, providing valuable insights for future exploration.
    \item The extrapolation capability of ML models are usually quite limited, i.e., they only performs well within the range of training dataset. In this regards, adopting transfer learning may remedy this issue to some extent. Transfer learning is a ML technique where a model trained on one task is reused or adapted for a different but related task. It leverages knowledge gained from one domain to another, potentially saving time and resources in training new models from scratch. By fine-tuning or retraining only a portion of the model on a smaller dataset relevant to the new task, one can adapt the pre-trained model to perform well on the new task. 
    ML models trained for a certain microstructure type may be adapted to similar microstructures by re-training with a much smaller dataset. 
    \item In the majority of existing works, the ML models are trained by data from simulations. Although it is time-consuming and expensive to collect experimental data, it may be beneficial to at least using a hybrid dataset \cite{wang2022data} consisting of both simulation and experiment data.
    \item Effective curation and sharing of both data and code are crucial for enabling practitioners to replicate, compare, reuse, and advance existing ML models, as well as to construct innovative models by leveraging available datasets. Consequently, this paves the way for the rapid and positive development of the field of ML-assisted modeling and design of microstructured materials.

\end{enumerate}
We anticipate that ML methods will become standard tools for addressing engineering challenges, like traditional numerical methods such as the FE method. Hopefully, this paper will contribute to paving the way for this trend.
\section*{Acknowledgements}
B.-X. Xu acknowledges the financial support of German Science Foundation (DFG) in the framework of the Collaborative Research Centre Transregio 270 (CRC-TRR 270, project number 405553726, sub-projects A06, B07, Z-INF) and 361 (CRC-TRR 361, project number 492661287, sub-projects A05), the Research Training Groups 2561 (GRK 2561, project number 413956820, sub-project A4), the Priority Program 2256 (SPP 2256, project number 441153493) and 2122 (SPP 2122, project number 493889809).  The authors also greatly appreciate the access to the Lichtenberg II High-Performance Computer (HPC) and the technique supports from the HHLR, Technische Universit\"at Darmstadt. The computating time on the HPC is granted by the NHR4CES Resource Allocation Board under the project ''special00007''.

\bibliographystyle{naturemag}

\end{document}